\documentclass[fleqn,usenatbib]{mnras}

\usepackage{newtxtext,newtxmath}
\usepackage[T1]{fontenc}
\usepackage{ae,aecompl}

\usepackage{graphicx}	% Including figure files
\usepackage{amsmath}	% Advanced maths commands
\usepackage{amssymb}	% Extra maths symbols

\usepackage{xcolor}
\definecolor{amber}{rgb}{1.0, 0.75, 0.0}

\definecolor{pink}{RGB}{255, 20, 147}

\definecolor{carminered}{rgb}{1.0, 0.0, 0.22}

\definecolor{byzantine}{rgb}{0.74, 0.2, 0.64}

\definecolor{amethyst}{rgb}{0.6, 0.4, 0.8}

\definecolor{blue-violet}{rgb}{0.54, 0.17, 0.89}

\definecolor{atomictangerine}{rgb}{1.0, 0.6, 0.4}

\definecolor{comment}{RGB}{166, 38, 164}

\title{Gravitational-wave follow-up with CTA after the detection of GRBs in the TeV energy domain}
\author[I. Bartos et al.]{%
I. Bartos,$^{1}$\thanks{E-mail: imrebartos@ufl.edu}, K.R. Corley$^2$, N. Gupte$^{1}$, N. Ash$^{1}$, Z. M\'arka$^2$, S. M\'arka$^2$ \\
$^1$Department of Physics, University of Florida, PO Box 118440, Gainesville, FL 32611-8440, USA\\
$^2$Department of Physics, Columbia University in the City of New York, 550 W 120th St., New York, NY 10027, USA\\
}

\begin{document}
\label{firstpage}
\pagerange{\pageref{firstpage}--\pageref{lastpage}}
\maketitle

\begin{abstract}
The recent discovery of TeV emission from gamma-ray bursts (GRBs) by the MAGIC and H.E.S.S. Cherenkov telescopes confirmed that emission from these transients can extend to very high energies. The TeV energy domain reaches the most sensitive band of the Cherenkov Telescope Array (CTA). This newly anticipated, improved sensitivity will enhance the prospects of gravitational-wave follow-up observations by CTA to probe particle acceleration and high-energy emission from binary black hole and neutron star mergers, and stellar core-collapse events. Here we discuss the implications of TeV emission on the most promising strategies of choice for the gravitational-wave follow-up effort for CTA and Cherenkov telescopes more broadly. We find that TeV emission (i) may allow more than an hour of delay between the gravitational-wave event and the start of CTA observations; (ii) enables the use of CTA's small size telescopes that have the largest field of view. We characterize the number of pointings needed to find a counterpart. (iii) We compute the annual follow-up time requirements and find that prioritization will be needed. (iv) Even a few telescopes could detect sufficiently nearby counterparts, raising the possibility of adding a handful of small-size or medium-size telescopes to the network at diverse geographic locations. (v) The continued operation of VERITAS/H.E.S.S./MAGIC would be a useful compliment to CTA's follow-up capabilities by increasing the sky area that can be rapidly covered, especially in the United States and Australia, in which the present network of gravitational-wave detectors is more sensitive.
\end{abstract}

\begin{keywords}
gravitational waves -- gamma-ray bursts
\end{keywords}

%%%%%%%%%%%%%%%%%%%%%%%%%%%%%%%%%%%%%%%%%%%%%%%%%%

%%%%%%%%%%%%%%%%%%%%%%%%%%%%%%%%%%%%%%%%%%%%%%%%%%%%%%%%%%%%%%%%%%%%%%%%%%%%%%%%%
\section{Introduction} \label{sec:introduction}
%%%%%%%%%%%%%%%%%%%%%%%%%%%%%%%%%%%%%%%%%%%%%%%%%%%%%%%%%%%%%%%%%%%%%%%%%%%%%%%%%

The multi-messenger discovery of the neutron star merger GW170817 demonstrated the promise of follow-up efforts to identify and interpret counterparts of gravitational-wave sources \citep{2017PhRvL.119p1101A,2017ApJ...848L..12A}. Observations by dozens of instruments identified the closest known GRB associated with the merger \citep{2017ApJ...848L..13A}, found that high-energy emission was detected at an unprecedented large viewing angle \citep{2017ApJ...848L..14G}, gave estimates on the expansion of the Universe \citep{2017Natur.551...85A}, established that the relativistic outflow was structured and likely interacted with the slower dynamical ejecta from the merger \citep{2018ApJ...856L..18M,2018PhRvL.120x1103L,2017Sci...358.1559K,2018Natur.561..355M,2019Sci...363..968G}, and constrained high-energy emission from the event \citep{2017ApJ...850L..35A,2018ApJ...861...85A,2018PhRvD..98d3020K}.

The Cherenkov Telescope Array (CTA) will soon join multi-messenger follow-up efforts with unprecedented sensitivity to sources producing very high-energy ($>$\,100 GeV) gamma-ray emission \citep{2019scta.book.....C}. CTA operations are expected to start with a partial array in 2022, while the operation of the full array is expected by 2025.

The beginning of CTA's operation will coincide with the vast expansion in gravitational-wave search capabilities \citep{2018LRR....21....3A}. Beyond the Advanced LIGO and Advanced Virgo observatories reaching their design sensitivity \citep{2015CQGra..32g4001L,2015CQGra..32b4001A}, they will be further upgraded by around 2025 to increase their monitored volume by almost an order of magnitude \citep{LIGOinstrumentwhitepaper}. In addition, the KAGRA gravitational wave detector and possibly LIGO-India will become operational and may reach their design sensitivity by this time, enhancing network sensitivity and improving on source localization \citep{PhysRevD.88.043007,LIGOIndia}. 

Gamma-ray bursts (GRBs), the main targets of gravitational-wave follow-up observations with CTA \citep{2014MNRAS.443..738B,2018MNRAS.477..639B,2018JCAP...05..056P,2019ICRC...36..790S}, are known to emit radiation beyond GeV energies. Gamma-ray emission from GRBs have been detected up to tens of GeV energies by the Large Area Telescope on the {\sl Fermi} satellite ({\sl Fermi}-LAT) \citep{2013ApJS..209...11A}, with no clear cutoff. These observations included short GRBs as well, such as GRB\,090510~\citep{2009Natur.462..331A,2010ApJ...716.1178A}. Short GRBs are associated with binary neutron star or neutron star--black hole mergers, which are primary sources of gravitational waves 
\citep{2001astro.ph.10349H,2003SPIE.4856..222M,2005PhRvD..72d2002A,2008ApJ...681.1419A,2008CQGra..25k4051A,2010JPhCS.243a2001M,2011GReGr..43..437C,2011CQGra..28k4013M,2013PhRvL.110x1101B,2013CQGra..30l3001B,2017ApJ...848L..12A}.
%These observations are also limited due to the universe being opaque at the highest energies for sources at typical GRB distances. 
Recently, gamma-ray emission up to TeV range has been discovered from three events: GRB\,190114C \citep{GRB190114C} by the MAGIC Telescope, GRB\,180720B by the H.E.S.S. Telescope \citep{GRB180720B}, and GRB\,190829A also by H.E.S.S. \citep{GRB190829A}.

These TeV discoveries substantially broaden the prospects of high-energy observations with CTA: (1) CTA is the most sensitive around $1\,$TeV \citep{2019APh...111...35A}, enabling deeper and/or more delayed follow-up observations. (2) all CTA telescopes are sensitive in the TeV band \citep{2019scta.book.....C}, enabling the use of even the Small Size Telescopes (SST; \citealt{2015arXiv150806472M}) in follow-up surveys. SSTs have greater field of view (FoV) and will be more numerous than CTA's other telescopes \citep{2019scta.book.....C}. (3) Even off-axis emission could be detectable with significantly increased coincident event rate, where photon energies are reduced \citep{2001ApJ...554L.163I}.

In this paper we examine the consequence of discernible TeV emission on the prospects of gravitational-wave follow-up observations with CTA. We first determine the expected high-energy luminosity from gravitational-wave sources that can be identified by CTA, and in particular the allowed delay for the start of follow-up observations following a gravitational-wave detection, which affects possible strategies (Section \ref{sec:sensitivity}). Second, we investigate the number of necessary pointings with CTA to cover expected gravitational-wave sky areas to understand the effect of localization uncertainty on optimal follow-up strategies (Section \ref{sec:pointing}). Third, we quantify the expected duration of follow-up observations needed from CTA to understand the feasibility of targeting different plausible source categories (Section \ref{sec:time}). Finally, we investigate whether and how a small number of SSTs by themselves could be used for follow-up (Section \ref{sec:partial}). 

%%%%%%%%%%%%%%%%%%%%%%%%%%%%%%%%%%%%%%%%%%%%%%%%%%%%%%%%%%%%%%%%%%%%%%%%%%%%%%%%%
\section{Detecting Short GRBs with CTA} \label{sec:sensitivity}
%%%%%%%%%%%%%%%%%%%%%%%%%%%%%%%%%%%%%%%%%%%%%%%%%%%%%%%%%%%%%%%%%%%%%%%%%%%%%%%%%

To determine the sensitivity of CTA to short GRBs, we consider the luminosity of GRB\,090510. {\sl Fermi}-LAT detected GeV emission from this GRB up to $\sim100$\,s following the burst \citep{2010ApJ...716.1178A}. This duration is much longer than the prompt MeV emission from the GRB, suggesting an alternative production mechanism. 

GeV emission from GRBs typically follows a power-law $t^{-\alpha}$ with temporal decay index $\alpha=0.99\pm0.8$ \citep{Ajello_2019}, consistent with an external shock afterglow origin (e.g., \citealt{2010MNRAS.403..926G,2010MNRAS.409..226K,2012RAA....12.1139M}). For GRB\,090510, we adopt a temporal decay index $\alpha=(2-3p)/4\approx1.38$ at MeV energies, where $p=2.5$ is the index of the power-law electron energy distribution. This index is consistent with observations \citep{2010ApJ...709L.146D}. 

For the energy index we will use $\beta\approx p/2=1.25$, which is the expected index for a forward shock origin for a power-law electron energy distribution with index $p=2.5$. This index is consistent with that found by \cite{2010ApJ...709L.146D}. We assume that emission follows this spectrum up to TeV energies. Our results are not sensitive to the cutoff energy as long as it is above $\sim1\,$TeV. With these, the spectral flux density of GRB\,090510 consistent with {\sl Fermi}-LAT measurements is \citep{2010ApJ...709L.146D}
\begin{equation}
F = 6\times10^{-8}\,\mu\mbox{Jy} \left(\frac{t}{100\,\mbox{s}}\right)^{-1.38}\left(\frac{E}{1\,\mbox{TeV}}\right)^{-1.25}\left(\frac{d_{\rm L}}{500\,\mbox{Mpc}}\right)^{-2}  
\end{equation}
where $t$ is time since the start of the GRB, $E$ is photon energy and $d_{\rm L}$ is the GRB's luminosity distance. 

To compare this flux spectrum to CTA's sensitivity, we calculate the expected flux in the energy band $[0.6\mbox{\,TeV},1\mbox{\,TeV}]$, and compare it to CTA's differential sensitivity\footnote{\url{https://www.cta-observatory.org/science/cta-performance/\#1525680063092-06388df6-d2af}. The choice of $[0.6\mbox{\,TeV},1\mbox{\,TeV}]$ energy band corresponds to a bin size of 5 per decade (see e.g. \citealt{2017APh....93...76H}).}. We scale the available differential sensitivity estimates at 250\,GeV to the energy band $[0.6\mbox{\,TeV},1\mbox{\,TeV}]$ using an improvement of a factor of 1.5, similar to the improvement expected for longer-duration observations (see, e.g., \citealt{2019APh...111...35A}). We conservatively omit the contribution of lower-energy emission, where CTA is less sensitive. 

We consider that CTA carries out consecutive exposures of uniform duration $t_{\rm exp}$ to cover the gravitational-wave localization region. We calculate results for two scenarios. In one we choose a fiducial distance of 500\,Mpc, which is the average distance at which a binary neutron star merger will be detectable with LIGO A+ for low-inclination sources \citep{LIGOinstrumentwhitepaper}. Inclination here refers to the angular distance between the line of sight and the binary orbital axis. At this distance TeV flux is attenuated by the cosmic microwave background by a factor of $\sim 0.4$ \citep{2009MNRAS.399.1694G}. For this distance scale we assume an exposure time of $t_{\rm exp}=100$\,s. In a second scenario we choose a source distance of 300\,Mpc, which is the expected distance range of Advanced LIGO/Virgo for binary neutron star mergers with low inclination \citep{2018LRR....21....3A}. There is negligible attenuation on this distance scale. For this scenario we adopt a shorter exposure time of $t_{\rm exp}=10$\,s.

\begin{figure*}%[htbp]
    \centering
    \includegraphics[width=0.5\linewidth]{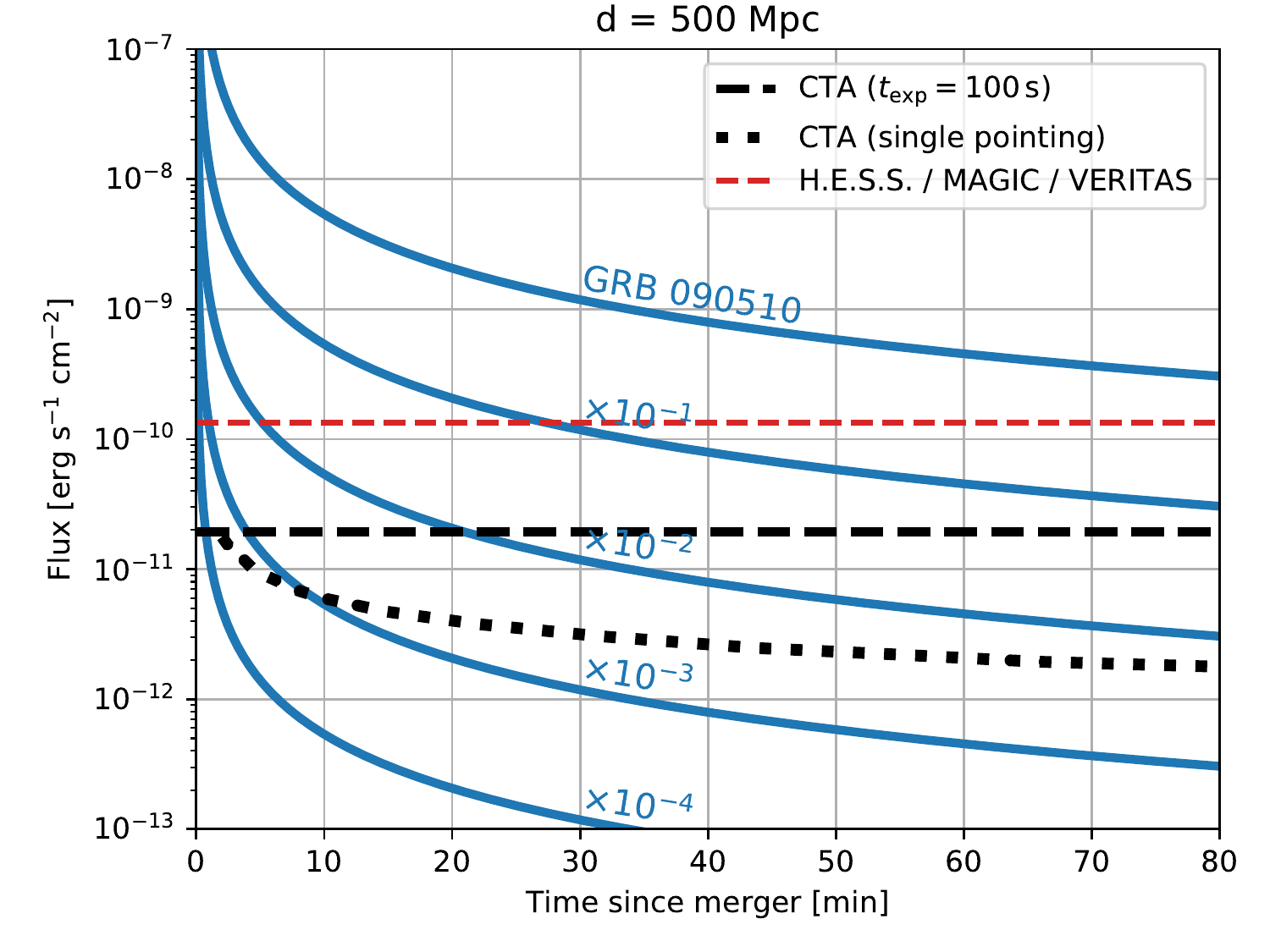}\includegraphics[width=0.5\linewidth]{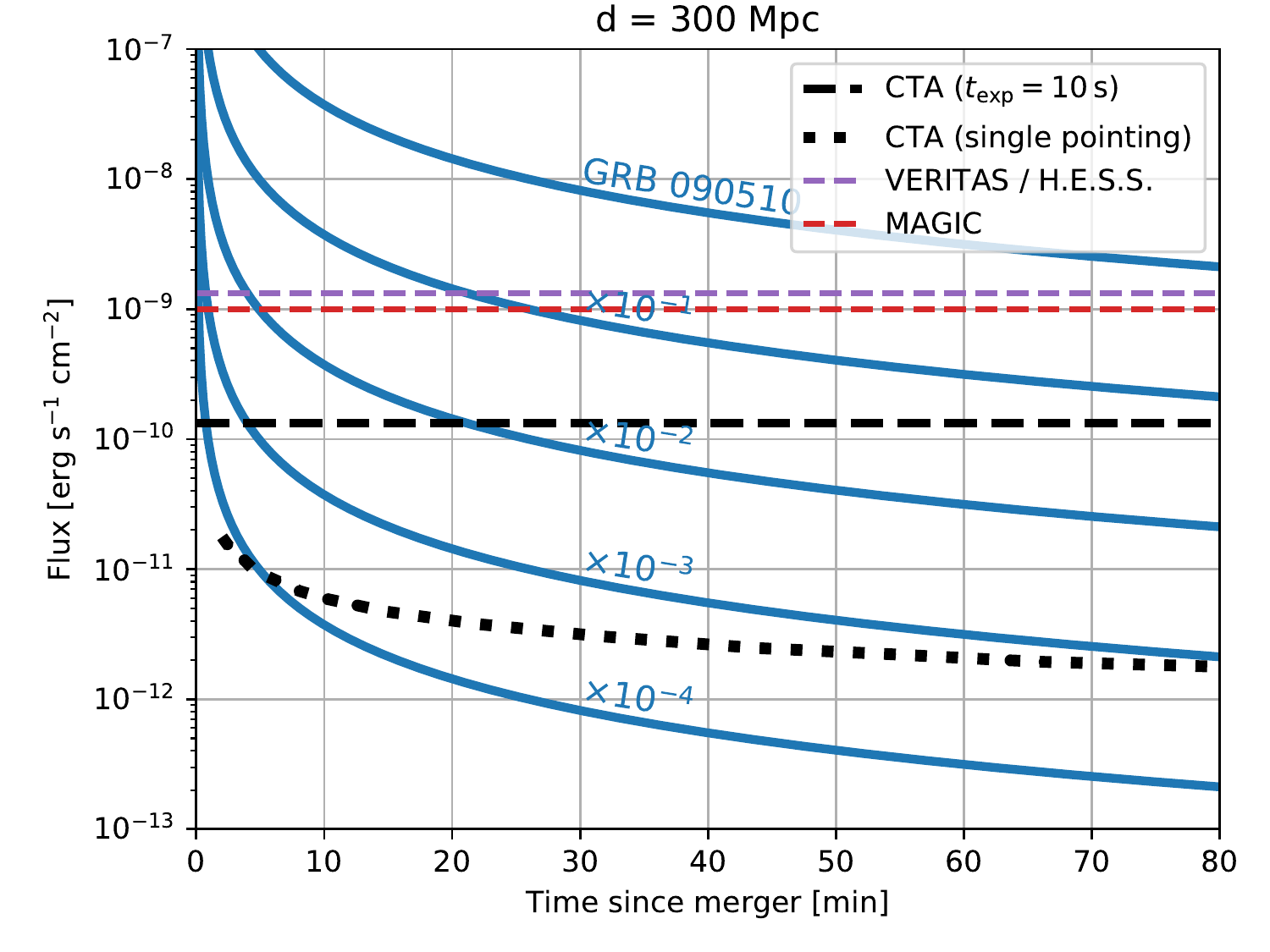}
    \caption{\label{fig:detectiontime} Expected TeV flux from GRB\,090510 at 500\,Mpc (left) and 300\,Mpc (right) and CTA's differential sensitivity for $t_{\rm exp}=100$\,s (left) and $t_{\rm exp}=10$\,s (right), as functions of time. Also shown are flux estimates from sources $1-4$ orders of magnitude weaker than GRB\,090510 (solid lines, see text in figures). CTA's differential sensitivity for continuous exposure with a single pointing is also shown (see legend). For comparison, we show the sensitivities of MAGIC (adopted from \citealt{DeividICRC}), VERITAS (adopted from \citealt{DeividICRC}) and H.E.S.S. (assumed to be the same as VERITAS).}
\end{figure*}

Fig. \ref{fig:detectiontime} shows the expected flux from a GRB\,090510-like event at 500\,Mpc (left) and 300\,Mpc (right) in comparison to CTA's differential sensitivity (for scenarios 1 and 2, respectively). We see that the event is easily detectable even hours after the GRB itself. Since GRB\,090510 is a particularly bright event, for comparison we also show the flux assuming $1-4$ orders of magnitude reduction. We see that even with such reduced emission, the event could be detectable by CTA if the the correct source direction is observed sufficiently fast. For example for a source $10^{-3}$ times weaker than GRB\,090510, TeV emission is detectable up to 5 minutes after the GRB.

We also consider the case of precisely known source direction, which could be due to a well localized gravitational-wave event (see Section \ref{sec:pointing}), the detection of the MeV emission by {\sl Fermi} or {\sl Swift}, or the coincident detection of high-energy neutrinos \citep{2019arXiv190105486C,2018arXiv181011467B,2012PhRvD..85j3004B,2011APh....35....1B,2011PhRvL.107y1101B}. In this case, exposure time can be significantly longer, which improves sensitivity. We show in Fig. \ref{fig:detectiontime} CTA's sensitivity as a function of exposure time. We see that for about 5 minutes exposure, sensitivity can improve by an order of magnitude compared to $t_{\rm exp}=10\,s$, making TeV sources $10^{-3}$ times weaker than GRB\,090510 detectable at 300\,Mpc even if observations start with a delay of $\sim30$\,min.

%=================================================================
%=================================================================

%%%%%%%%%%%%%%%%%%%%%%%%%%%%%%%%%%%%%%%%%%%%%%%%%%%%%%%%%%%%%%%%%%%%%%%%%%%%%%%%%
\section{How many CTA pointings survey a gravitational-wave skymap?} \label{sec:pointing}
%%%%%%%%%%%%%%%%%%%%%%%%%%%%%%%%%%%%%%%%%%%%%%%%%%%%%%%%%%%%%%%%%%%%%%%%%%%%%%%%%

The number of CTA pointings needed to survey the whole gravitational-wave localization sky area (skymap) affects the possible depth of CTA's follow-up observations as well as the required telescope time. In order to quantify this number, we simulated the detection of 500 neutron-star mergers via gravitational waves.

\begin{figure*}%[htbp]
    \centering
    \includegraphics[width=0.5\linewidth]{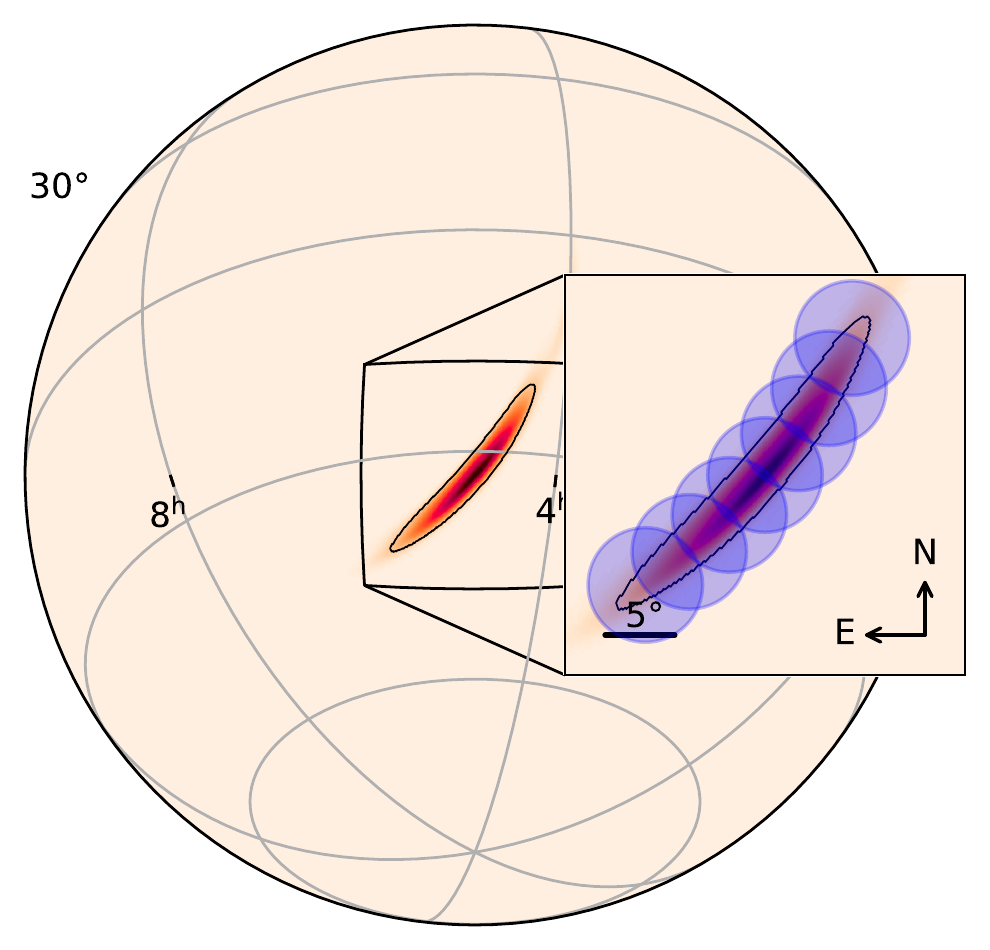}\includegraphics[width=0.5\linewidth]{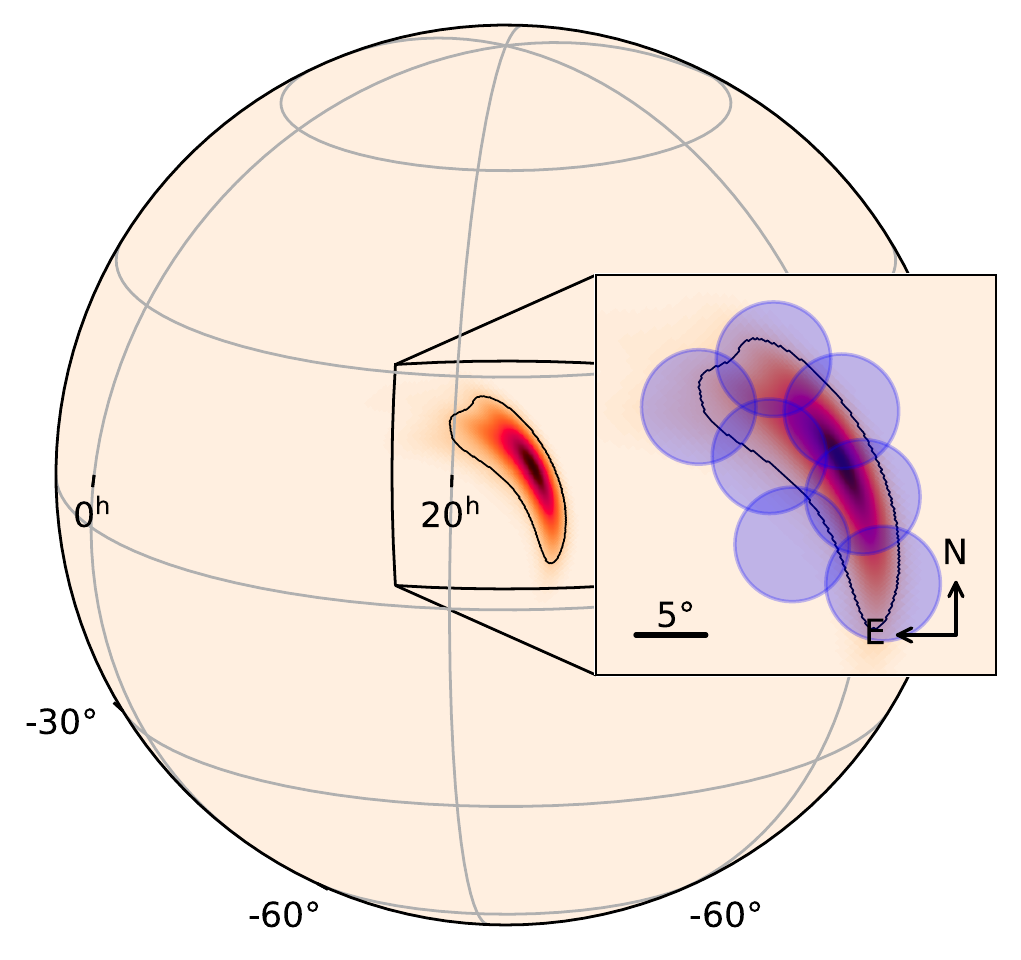}
    \caption{\label{fig:skymapexample} Example gravitational-wave skymaps (90\% credible level) for binary neutron star mergers, reconstructed by BAYESTAR. Also shown are CTA tiling pattern suggested by the 'greedy' method (left) and 'honeycomb' method (right).}
\end{figure*}

\begin{figure}%[htbp]
    \centering
    \includegraphics[width=\linewidth]{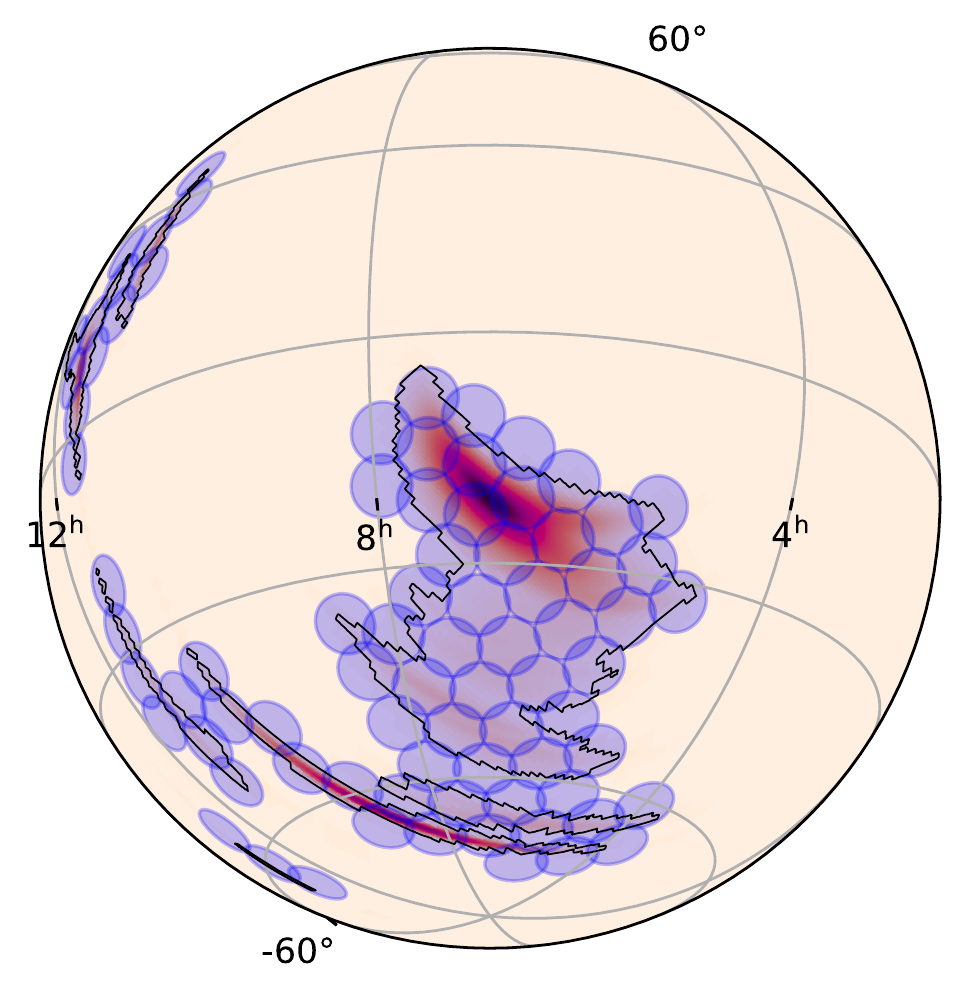}
    \caption{\label{fig:skymapexample2} An example for a poorly localized gravitational-wave skymap (90\% credible level) for binary neutron star mergers, reconstructed by BAYESTAR. Also shown are CTA tiling pattern using the 'honeycomb' method.}
\end{figure}

For each neutron star merger we randomly selected a location, uniformly distributed in volume around Earth. We adopted advanced LIGO and advanced Virgo design sensitivities to determine for each event whether it is detectable, requiring for detection a network signal-to-noise ratio (SNR) of 12 and at least two detector SNRs of 4. We separately considered gravitational-wave events detected by only two detectors (``2-detector" case) and detected by both LIGO detectors and the Virgo detector (``3-detector" case). For each detected event, we reconstructed its localization skymap using the BAYESTAR algorithm \citep{2016PhRvD..93b4013S}. Example localizations are shown in Figs. \ref{fig:skymapexample} \& \ref{fig:skymapexample2}. 

We then constructed CTA tilings that cover the gravitational-wave skymaps. We use tiles with $8^\circ$ diameter, reflecting the field-of-view of CTA's SSTs \citep{2019scta.book.....C}. We tested two different tiling methods. The first, called {\it ``greedy"} method iteratively selects the sky location with the highest gravitational-wave probability density from the sky locations not already covered by a tile (see Fig. \ref{fig:skymapexample} for an example). The second, a {\it ``honeycomb"} method, requires the tiles to be organized in a hexagonal structure, and optimizes for the location and orientation of the hexagon to minimize the number of required tilings (see Fig. \ref{fig:skymapexample2} for an example). This latter method is similar to some of the CTA tiling strategies suggested in the literature \citep{2013APh....43..317D,2018JCAP...05..056P}. We find that the two methods perform similarly for most skymaps, except for the largest ones above $\gtrsim100$\,deg$^{2}$ for which the ``honeycomb" method is better, i.e. it can cover the skymap with less tiles. For the 3-detector case, the median number of required tiles is 4 and 3 for the {\it ``greedy"} and {\it ``honeycomb"} methods, respectively, while for the 2-detector case the median number of required tiles are 28 and 23 for the two methods, respectively.

%=================================================================
%=================================================================
%=================================================================
%=================================================================

\begin{figure}%[htbp]
    \centering
    \includegraphics[width=\linewidth]{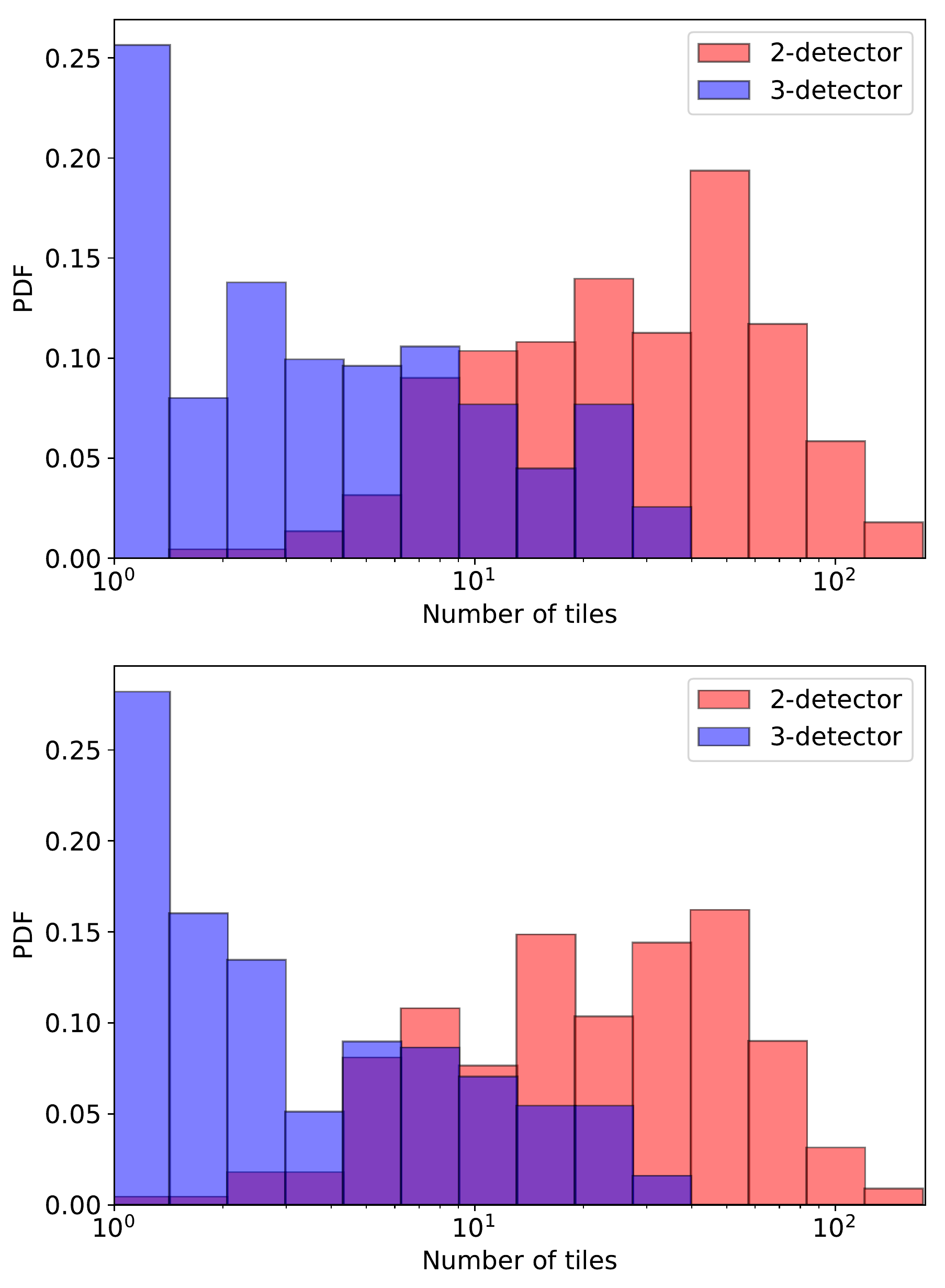}
    \caption{\label{fig:tilenum} Probability density function (PDF) of the number of CTA tilings needed to cover gravitational-wave skymaps of neutron star mergers, separately for the 2-detector and 3-detector cases (see legend). We assumed that LIGO and Virgo are at their design sensitivities. Results here are shown for the 'greedy' (top) and 'honeycomb' (bottom) tiling methods, which give similar results.}
\end{figure}

The obtained probability distribution of the number of CTA pointings needed to cover the gravitational-wave skymaps (at 90\% credible level) is shown in Fig. \ref{fig:tilenum}. Interestingly, we find that about 50\% of skymaps for the 3-detector case can be covered by a single CTA pointing, therefore making it feasible for a significant fraction of gravitational-wave detections to carry out deep follow-up observations with CTA. In general, we see that most 3-detector skymaps can be covered with $1-10$ pointings, while most 2-detector skymaps need $10-100$ pointings. 

%=================================================================
%=================================================================

%%%%%%%%%%%%%%%%%%%%%%%%%%%%%%%%%%%%%%%%%%%%%%%%%%%%%%%%%%%%%%%%%%%%%%%%%%%%%%%%%
\section{Follow-up time needed} \label{sec:time}
%%%%%%%%%%%%%%%%%%%%%%%%%%%%%%%%%%%%%%%%%%%%%%%%%%%%%%%%%%%%%%%%%%%%%%%%%%%%%%%%%

Here we estimate the total required follow-up time expected for different gravitational-wave source classes. The purpose of this exercise is to understand whether all gravitational-wave detections can be comprehensively followed up by CTA or whether some level of optimization or prioritization is needed.

Taking into account that gravitational-wave detectors are in observing mode in about $\sim 70\%$ of the time \citep{2018LRR....21....3A}, about 45\% of events will be found by three detectors and 55\% by two, since we count any two detectors out of the three. Here for simplicity we ignore cases in which a single detector discover an event, as well as the different distance ranges of the 2-detector and 3-detector cases. Assuming that typically 10 pointings are needed to cover the skymap for the 3-detector case and 100 pointings for the 2-detector case, we find an average number of 60 pointings per event. 

To calculate the CTA telescope time needed per event, we will adopt an initial CTA slew time of $t_{\rm slew, initial}=20$\,s, then a slew time of $t_{\rm slew}=5$\,s between pointings. For an exposure time of $t_{\rm exp}=10$\,s, this corresponds to an average 15\,minutes of CTA time needed for follow-up per event, while for $t_{\rm exp}=100$\,s it is 1.8\,hours per event. 

Given the currently planned CTA follow-up observation time of 5\,hrs per year per site once construction is completed, it is clear that it will not be possible to follow up all gravitational-wave events; prioritization will be needed. 

The neutron star merger detection rate during LIGO A+'s operation can be estimated considering LIGO A+'s detection range of 325\,Mpc and the merger rate of $110-3840$\,Gpc$^{-3}$yr$^{-1}$ \citep{2018arXiv181112907T}, giving an expected detection rate of $20-600$\,yr$^{-1}$. Considering CTA's expected duty cycle of $15$\%, this corresponds to a total possible follow-up rate of $3-90$\,yr$^{-1}$.

The most straightforward prioritization is to require that an event is detected with three gravitational-wave detectors. This puts the number of required pointings to $\lesssim 10$, while the number of events drops by a factor of 3. For $t_{\rm exp}=100$\,s exposures, the localization area of such events can be covered within 15\,min, making the total required time $0.2-7$\,hrs per year. If a total 5+5\,hrs of follow-up time for the two CTA sites, this means that all neutron star mergers detected by three detectors will be possible to follow up.

%One can also separately consider more nearby events, within 300\,Mpc, which require less exposure time. The number of neutron star mergers within 300\,Mpc is $7-220$\,yr$^{-1}$, corresponding to a total possible CTA follow-up rate of $1-30$\,yr$^{-1}$. Taking $t_{\rm exp}=10$\,s for these events (see Fig. \ref{fig:detectiontime}), and 100 required pointings (see Fig. \ref{fig:tilenum}), the required CTA obseving time is 25\,min, making the worst localized events unfeasible to follow up. 

Binary black hole mergers represent a less certain but potentially more disruptive follow-up opportunity. While most formation channels predict no electromagnetic counterpart, binary black holes merging in dense gaseous environments may accrete matter and may produce high-energy emission, e.g., in the disks of active galactic nuclei \citep{2017ApJ...835..165B,2017NatCo...8..831B,2019arXiv190703746M,2012MNRAS.425..460M,2017MNRAS.464..946S,2019arXiv190609281Y,2019ApJ...876..122Y,2019arXiv190202797C}. The distribution of localization uncertainty for black hole mergers is similar to that for neutron star mergers \citep{2018LRR....21....3A}, therefore we will adopt the same estimated follow-up time per event. The main difference is the detection rate of black hole mergers. To estimate this rate we adopt a detection range of 2.5\,Gpc for LIGO A+ for a fiducial 30\,M$_\odot-30$\,M$_\odot$ binary black hole merger \citep{LIGOinstrumentwhitepaper} and consider a black hole merger rate of $9.7-101$\,Gpc$^{-3}$yr$^{-1}$ \citep{2018arXiv181112907T}, obtaining a detection rate of $600-6000$\,yr$^{-1}$. An added uncertainty we neglect here is that the black hole merger detection range roughly linearly depends on the black hole masses, therefore the overall detection rate depends on the currently uncertain black hole mass distribution. Even after reducing this rate with CTA's 15\% duty cycle, the corresponding required total follow-up time is prohibitive. We conclude that for black hole mergers it will be necessary to significantly downselect targets for CTA follow-up. Prioritization can be achieved, e.g., by considering only the events with the most accurate sky localization. For example, focusing only on events whose localization can be covered by a single CTA pointing still leaves a follow-up rate of $10-100$\,yr. Assuming $t_{\rm exp}=100$\,s, this is $0.3-3$\,hrs per year, which could be a feasible addition to the follow-up program.

Another possible downselection criterion can be the mass and spin of the black holes, with higher masses and spins expected from dynamical formation channels such as in the disks of active galactic nuclei \citep{2019ApJ...876..122Y,2019arXiv190609281Y}. Note, however, that at this point LIGO/Virgo do not publicly disclose this information prior to publication.

There are additional source types that will affect the needed overall follow-up rate. These include neutron star--black hole mergers; the rate of such mergers is currently uncertain but their detection rate is lower than the previous two event types. Another general event type of possible interest is weak, not confidently established gravitational-wave event candidates. The rate of these, so-called sub-threshold events can be arbitrarily set using a desired significance threshold, in order to balance high false alarm rate and low false dismissal rate.

%%%%%%%%%%%%%%%%%%%%%%%%%%%%%%%%%%%%%%%%%%%%%%%%%%%%%%%%%%%%%%%%%%%%%%%%%%%%%%%%%
\section{Can incomplete array configurations be useful?} \label{sec:partial}
%%%%%%%%%%%%%%%%%%%%%%%%%%%%%%%%%%%%%%%%%%%%%%%%%%%%%%%%%%%%%%%%%%%%%%%%%%%%%%%%%

CTA's high sensitivity to the high-energy transients considered here means that a partially completed detector is also useful in follow-up surveys, and should be utilized even before the array is complete. Additionally, especially for sources extending to the TeV energy domain, even a limited number of SSTs could be sufficient to probe high-energy emission. SSTs cost significantly less than CTA's mid-sized or large-sized telescopes \citep{2019scta.book.....C}, potentially allowing more specialized and distributed deployment.

\begin{figure*}%[htbp]
    \centering
    \includegraphics[width=0.5\linewidth]{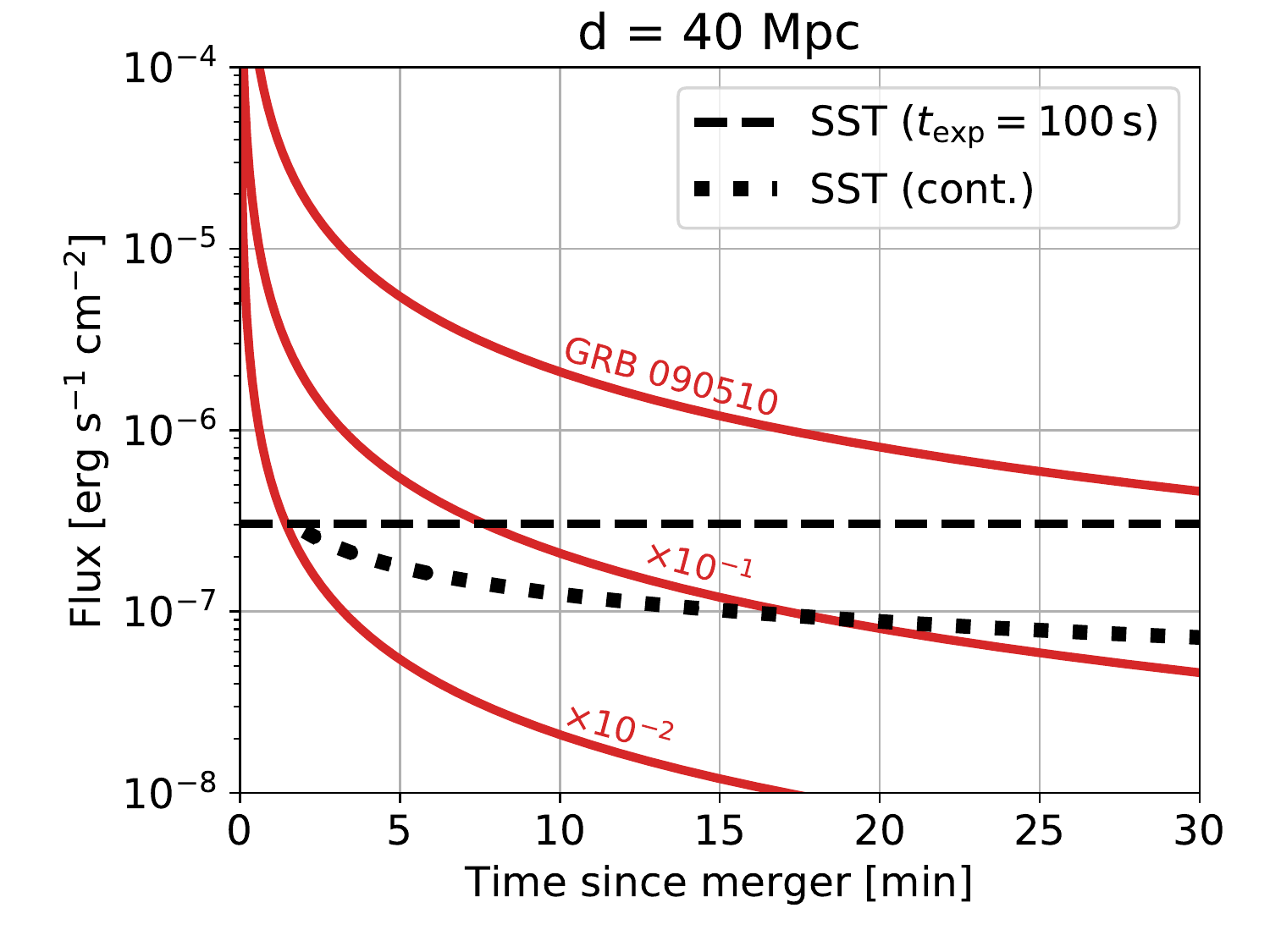}\includegraphics[width=0.5\linewidth]{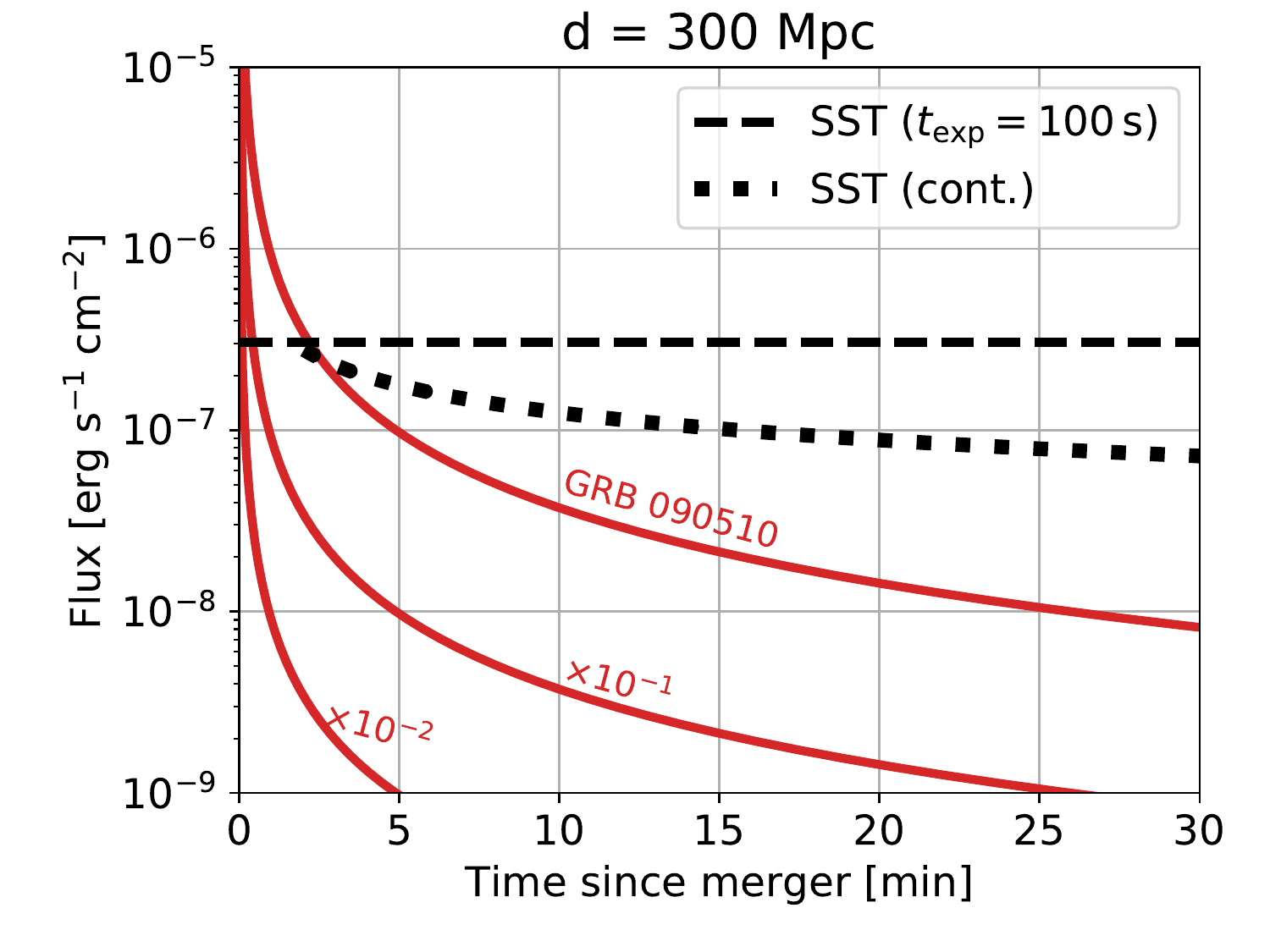} 
    \caption{\label{fig:detectiontimeFACT} Differential sensitivity of FACT, a single-SST telescope as a function of time, compared to the expected TeV flux from GRB\,090510 at 40\,Mpc (i.e. the distance of GW170817; left) and 300\,Mpc (right). Also shown are flux estimates from sources $1-2$ orders of magnitude weaker than GRB\,090510 (solid lines, see text in Figure). The SST's differential sensitivity for exposure time $t_{\rm exp}=100$\,s and continuous exposure are both shown (see legend). }
\end{figure*}

To characterize the potential of SSTs in following up gravitational-wave events, we consider the differential sensitivity of a single SST, obtained by the First G-APD Cherenkov Telescope (FACT; \citealt{2017ICRC...35..791N}). We use the same short-GRB emission model discussed in Section \ref{sec:sensitivity}, but this time consider a nearby event at 40\,Mpc, similar to the distance of GW170817 \citep{2017PhRvL.119p1101A}. The expected flux from an event like GRB\,090510 at a distance of 40\,Mpc, as well as the flux detection threshold of a single SST for $t_{\rm exp}=100$\,s exposure time, are shown in Fig. \ref{fig:detectiontimeFACT}. We see that assuming fast CTA response, even events that are $10^{-2}$ times fainter than GRB\,090510 would be at 40\,Mpc at 1\,TeV could be detected. While such nearby events that are sufficiently on-axis to produce detectable TeV emission are rare, they may occur over a period of several years, especially considering that several SSTs in unison could significantly expand the distance within which such events could be detected. In Fig. \ref{fig:detectiontimeFACT} we also compare FACT's sensitivity with a more distant GRB at 300\,Mpc. We see that at this, more typical distance a GRB\,090510-like event could be detected with sufficiently rapid ($\lesssim1$\,min) response, however fainter events would be difficult to detect.

Beyond supporting early surveys with an incomplete CTA, this also means that it may be beneficial to deploy a small number of SSTs farther from CTA's main northern and southern locations, e.g., worldwide. Such SSTs could significantly increase the sky coverage in the TeV energies, increasing the probability that a nearby gravitational wave event can be followed up.
%\Szabi{[Can we write a paper from SST location optimizations, by the way?]}

%=================================================================
%=================================================================

\begin{figure*}%[htbp]
    \centering
    \includegraphics[width=\linewidth]{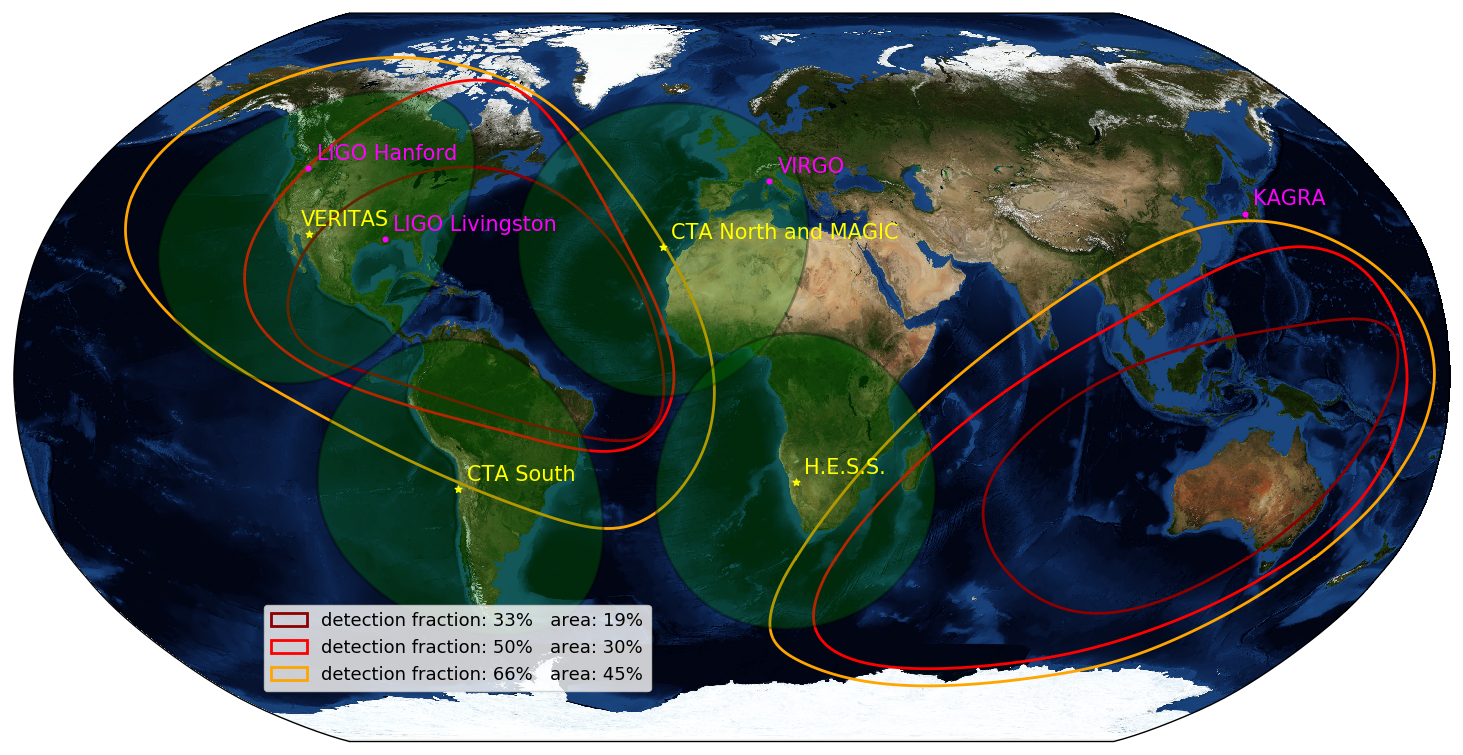}
    \caption{\label{fig:ligomap} Probability contours of the source directions of simulated gravitational wave discoveries by the LIGO/Virgo network, assuming uniform source distribution in co-moving volume (see legend). Directions above and 'below' the United states are preferred due to the sensitivity of the LIGO detectors. Also shown are the locations of gravitational-wave detectors and Cherenkov telescopes, and the sky areas the different Cherenkov telescopes can monitor at a given time, assuming a maximum zenith of 35$^\circ$.}
\end{figure*}

In addition to the immediate CTA response independent of location and weather and broadening the range on the sky that is surveyable at any moment at TeV energies, another factor in additional SST deployment could be the uneven directional sensitivity of gravitational wave detectors. Considering the LIGO and Virgo detectors at their design sensitivity, the detector network is the most sensitive roughly above and below the two LIGO detectors. In Fig. \ref{fig:ligomap} we demonstrate this uneven sensitivity by showing the sky areas corresponding to 33\%, 50\% and 66\% of the expected detection rate. We assumed LIGO and Virgo design sensitivities \citep{2018LRR....21....3A}. We see that half the events will be detected on a sky area less than a third of the sky, making the TeV coverage of these areas particularly important. We find similar contours if we exclusively consider more distant events. This could be a useful guide in considering additional SST deployment locations.

%=================================================================
%=================================================================

%%%%%%%%%%%%%%%%%%%%%%%%%%%%%%%%%%%%%%%%%%%%%%%%%%%%%%%%%%%%%%%%%%%%%%%%%%%%%%%%%
\section{Will currently existing Cherenkov telescopes be helpful?} \label{sec:otherobservatories}
%%%%%%%%%%%%%%%%%%%%%%%%%%%%%%%%%%%%%%%%%%%%%%%%%%%%%%%%%%%%%%%%%%%%%%%%%%%%%%%%%

While CTA will be the most sensitive ground-based gamma-ray detector, here we consider whether currently existing observatories, including H.E.S.S. \citep{2004NewAR..48..331H}, MAGIC \citep{2003NuPhS.114..247B} and VERITAS \citep{2008AIPC.1085..657H}, will be beneficial even after CTA's construction is completed. 

In Fig. \ref{fig:detectiontime} we show the differential sensitivity for H.E.S.S., MAGIC and VERITAS assuming $t_{\rm exp}=100\,s$ in comparison to the expected flux of a GRB\,090510-like event at 500\,Mpc distance, and assuming $t_{\rm exp}=10\,s$ in comparison for the same event at 300\,Mpc. We find that, while these detectors are approximately an order of magnitude less sensitive than CTA will be, they can still easily detect a GRB\,090510-like event at distances relevant to gravitational-wave observations. For sufficiently rapid follow-up even a 100 times weaker event would also be detectable. 

Currently existing facilities can be highly beneficial especially if they cover sky locations inaccessible to CTA. As CTA's duty cycle is 15\%, detectors located at complementary positions on Earth covering different directions can provide broader coverage similarly to distributed SSTs discussed in the previous section.

%%%%%%%%%%%%%%%%%%%%%%%%%%%%%%%%%%%%%%%%%%%%%%%%%%%%%%%%%%%%%%%%%%%%%%%%%%%%%%%%%
%\section{Off Axis} \label{sec:offaxis}
%%%%%%%%%%%%%%%%%%%%%%%%%%%%%%%%%%%%%%%%%%%%%%%%%%%%%%%%%%%%%%%%%%%%%%%%%%%%%%%%%

%%%%%%%%%%%%%%%%%%%%%%%%%%%%%%%%%%%%%%%%%%%%%%%%%%%%%%%%%%%%%%%%%%%%%%%%%%%%%%%%%
\section{Conclusion} \label{sec:conclusion}
%%%%%%%%%%%%%%%%%%%%%%%%%%%%%%%%%%%%%%%%%%%%%%%%%%%%%%%%%%%%%%%%%%%%%%%%%%%%%%%%%

We investigated the prospects of gravitational-wave follow-up observations with CTA in light of the recent discovery of TeV emission from GRBs. If such TeV emission is typical, CTA will have ample opportunity to discover very-high-energy gamma rays from gravitational-wave sources. This also opens up the possibility for surveys with the partially completed CTA, as well as the motivation to consider the expansion of the CTA array and the continued operation of current Cherenkov facilities. Our conclusions are the following:
\begin{itemize}
\item CTA could detect very-high-energy emission from short GRBs even if it starts observing with significant delay following the gravitational-wave detection of a neutron star merger. For bright emission expected from a GRB\,090510-like event this can be over an hour, while even for a $10^{-2}$ times fainter event 20\,min delay is acceptable (see Fig. \ref{fig:detectiontime}).
\item About 25\% of mergers detected with 3 gravitational-wave detectors will be sufficiently well-localized to have a single SST pointing cover the full 90\% C.L. gravitational-wave skymap. Most mergers found with 3 detectors require $1-10$ pointings, while mergers found when only two detectors are operational will require $10-100$ pointings to follow-up (see Fig. \ref{fig:tilenum}).
\item Following up all neutron star mergers detected by a network of three gravitational-wave detectors (LIGO/Virgo A+) will require $0.2-7$\,hrs per year of CTA time, while following up events detected while only two detectors were operational would be unfeasible. The follow-up of binary black hole mergers will also require significant prioritization. Focusing on binary black hole mergers whose skymap can be covered with a single CTA pointing will still represent $10-100$ events per year, corresponding to a feasible follow-up time of $0.3-3$\,hrs per year. Another possibility is to focus on the heaviest/spinning black holes most promising for producing electromagnetic emission.
\item Even a few SSTs or MSTs could be sufficient to probe TeV emission from gravitational-wave sources. Deploying such SSTs or MSTs farther from the main CTA sites could enhance the chance of detecting TeV emission from events, especially by covering sky locations over the United States and Australia due to the direction-dependent sensitivity of gravitational-wave detectors and their day-night sensitivity cycles. It will be useful to investigate the specific locations of an optimized distribution, and possible optimizations of relevant detector design.
\item Presently operational Cherenkov telescopes H.E.S.S. and VERITAS represent complementary sky coverage to CTA and are sufficiently sensitive to detect GRBs $10^{-2}$ times fainter than GRB\,090510 at the relevant 500\,Mpc, assuming that they can rapidly respond to multi-messenger triggers. Their continued operation could therefore meaningfully increase the rate of follow-up discoveries of TeV emission from gravitational-wave sources.
\end{itemize}

\section*{Acknowledgments}
The authors thank Tristano Di Girolamo, Brian Humensky, Chris Messenger, Reshmi Mukherjee and Deivid Ribeiro for useful discussions. The authors are thankful for the generous support of the University of Florida and Columbia University in the City of New York. The Columbia Experimental Gravity group is grateful for the generous support of the National Science Foundation under grant PHY-1708028. 

%%%%%%%%%%%%%%%%%%%%%%%%%%%%%%%%%%%%%%%%%%%%%%%%%%



\bibliographystyle{mnras}
\bibliography{Refs} % if your bibtex file is called example.bib

% \begin{thebibliography}{99}
% \bibitem[\protect\citeauthoryear{Author}{2012}]{Author2012}
% Author A.~N., 2013, Journal of Improbable Astronomy, 1, 1
% \bibitem[\protect\citeauthoryear{Others}{2013}]{Others2013}
% Others S., 2012, Journal of Interesting Stuff, 17, 198
% \end{thebibliography}

%%%%%%%%%%%%%%%%%%%%%%%%%%%%%%%%%%%%%%%%%%%%%%%%%%

%%%%%%%%%%%%%%%%% APPENDICES %%%%%%%%%%%%%%%%%%%%%

% \appendix

% \section{Some extra material}

% If you want to present additional material which would interrupt the flow of the main paper,
% it can be placed in an Appendix which appears after the list of references.

%%%%%%%%%%%%%%%%%%%%%%%%%%%%%%%%%%%%%%%%%%%%%%%%%%

% Don't change these lines
\bsp	% typesetting comment
\label{lastpage}
\end{document}